# Single-Photon Imager Based on Microwave Plasmonic Superconducting Nanowire

**Authors:** Qing-Yuan Zhao[1], Di Zhu[1], Niccolò Calandri[1,2], Andrew E. Dane[1], Adam N. McCaughan[1], Francesco Bellei[1], Hao-Zhu Wang[1], Daniel F. Santavicca[3], Karl K. Berggren[1*]

**Affiliations:**
[1]Massachusetts Institute of Technology, Department of Electrical Engineering and Computer Science, Cambridge, MA, 02139

[2]Politecnico di Milano, Department of Electronics, Information and Bioengineering, Milano, ITA, 22020

[3]University of North Florida, Department of Physics, Jacksonville, FL 32224

*Correspondence to: berggren@mit.edu

**Abstract**: Detecting spatial and temporal information of individual photons by using single-photon-detector (SPD) arrays is critical to applications in spectroscopy, communication, biological imaging, astronomical observation, and quantum-information processing. Among the current SPDs[1], detectors based on superconducting nanowires have outstanding performance[2], but are limited in their ability to be integrated into large scale arrays due to the engineering difficulty of high-bandwidth cryogenic electronic readout[3–8]. Here, we address this problem by demonstrating a scalable single-photon imager using a single continuous photon-sensitive superconducting nanowire microwave-plasmon transmission line. By appropriately designing the nanowire's local electromagnetic environment so that the nanowire guides microwave plasmons, the propagating voltages signals generated by a photon-detection event were slowed down to ~ 2% of the speed of light. As a result, the time difference between arrivals of the signals at the two ends of the nanowire naturally encoded the position and time of absorption of the photon. Thus, with only two readout lines, we demonstrated that a 19.7-mm-long nanowire meandered across an area of 286 μm × 193 μm was capable of resolving ~ 590 effective pixels while simultaneously recording the arrival times of photons with a temporal resolution of 50 ps. The nanowire imager presents a scalable approach to realizing high-resolution photon imaging in time and space.

**Main Text:**

Quantum and classical optics are currently limited by our ability to efficiently sense and process information about single photons. For example, to enhance the information-carrying capacity of a quantum channel[9] and improve security in quantum key distribution[10,11], information is typically encoded in the position and arrival time of individual photons. Determining the spatial and temporal information of photons is currently accomplished by single-photon detector (SPD) arrays. Among existing SPD array technologies, the transition edge sensor (TES) and the microwave kinetic inductance detector (MKID) provide moderate spectral information but less impressive temporal resolution (*e.g.*, the timing jitter is measured in nanoseconds for TESs[12] and microseconds for MKIDs[13]). Photomultiplier tubes and single-photon avalanche diodes have sub-1-ns timing jitter in the visible domain, but their detection performance deteriorates in the infrared, and scaling these technologies to large spatial arrays is challenging[1]. Improved timing performance of sub-20-ps timing jitter[14] and sub-10-ns recovery time[15] is possible with superconducting-nanowire single-photon detectors (SNSPDs), which also have been demonstrated to have near-unity detection efficiency[2], less than 1 dark-count per second (cps)[16], a wide spectral response from the visible to infrared[17] and greater than 100 cps counting rate[18]. However, attempts to create arrays of SNSPDs have had limited success[3–8]. Traditional row-column rectangular pixel arrays require large numbers of readout lines[5], while attempts at time-based and frequency-based multiplexing architectures require additional components within and between pixels, and thus suffer from low fill factors[6,8]. As a result, the current state-of-the-art SNSPD array is limited to ~ 100 pixels[5].

Typical operation of an SNSPD lacks spatial sensitivity, although the excitation due to the absorption of a photon is localized within the nanowire. In the simplest broadly-accepted detection model of an SNSPD[19], an absorbed photon generates a localized region of elevated electron temperature, *i.e.* a hotspot, which leads to the formation of a resistive domain across the nanowire. Unfortunately, the conventional electrical readout of an SNSPD could not be used to determine the location of the resistive domain because electrically the nanowire was modeled as a lumped-element inductor in series with a non-linear dynamic resistor that represented the resistive domain[20]. In such a scheme, the relative location of the resistor and inductor cannot change the output behavior of the circuit. Despite its wide popularity, a lumped-element model cannot completely describe the electrical behavior of nanowire detectors because nanowires in these detectors are typically longer than the wavelength of the electrical signals they carry. In such situations, a distributed-element model must be used [21,22].

Here, we describe an experiment in which we used a superconducting nanowire as the center conductor in a microwave plasmonic coplanar transmission line to determine the position of the hot-spot along the nanowire while preserving information about the time of arrival of the photon. This approach uses the photon-sensitive nanowire to realize a scalable single-photon imager, which we refer to as a superconducting nanowire single-photon imager (SNSPI) to distinguish it from an SNSPD. Unlike a transmission line made from normal metals, in the SNSPI transmission line, significant kinetic energy is stored in the motion of Cooper pairs. As a result, microwave plasmons propagate at a speed of ~ 2% of the speed of light in this medium, so that a relatively small propagation distance can result in a measurable time delay[21,23]. In addition, the guided plasmonic TM mode in the superconducting waveguide squeezes the field so that it is confined to a region ~ 200-nm in radius around the nanowire[24]. Finally, the transmission lines are terminated in Klopfenstein tapers to permit efficient coupling of the signal from the high-impedance transmission line to 50 Ω room-temperature amplifiers[25]. By using this approach, the electrical pulses triggered by photon absorption event were guided along the nanowire, enabling us to extract the time and position of the absorbed photon by using the relative arrival times of the output electrical pulses at the two ends of the nanowire.

The operation of the SNSPI is shown in Fig. 1a. After a photon is absorbed at location $x_p$ at time $t_p$, the increase of the local resistance generates two electrical pulses of opposite polarities, which propagate with a constant velocity $v$ towards the two ends of the transmission line, introducing delays $\tau_1$ and $\tau_2$, respectively. After removing the fixed delays from tapers, cables and readout electronics and extracting $\tau_1$ and $\tau_2$ from the arrival times of the electrical pulses, every ($x_p$, $t_p$) pair can be determined by using two linear functions: $x_p = ((\tau_2 - \tau_1)v + L)/2$ and $t_p = ((\tau_2 + \tau_1) - L/v)/2$. Such a delay-based readout is conceptually similar to the readout of a time-multiplexed SNSPD array[6]. However, the SNSPI simultaneously acts as the detector and a delay component, without any multiplexing circuits or clock signals, resulting in a dramatically more compact design that is suitable for large-scale integration.

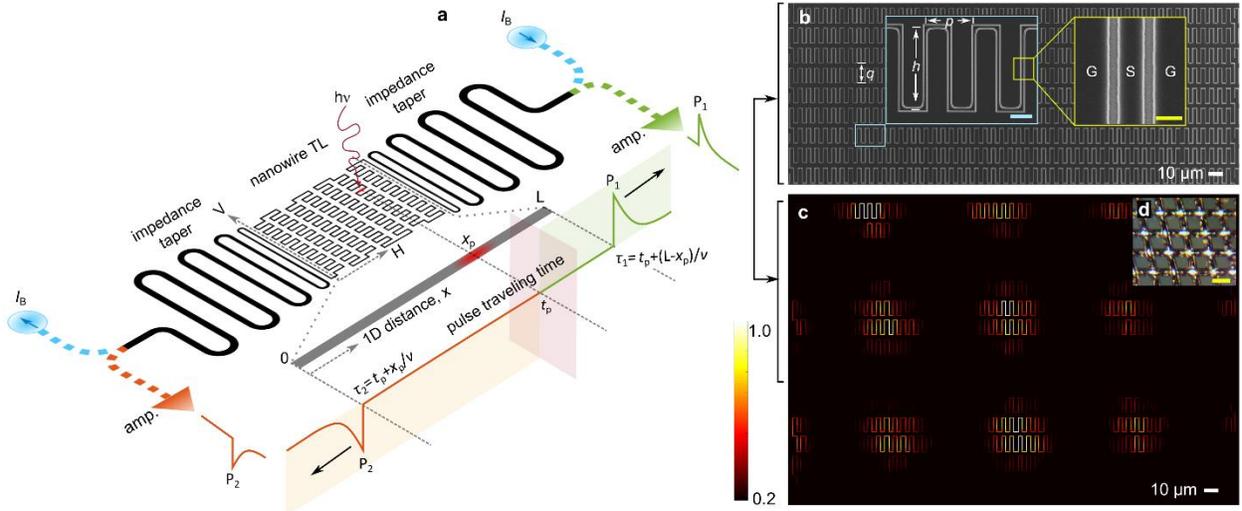

**Figure 1 | Superconducting nanowire single-photon imager (SNSPI). a**, Architecture of the SNSPI. The nanowire transmission line (TL) and the impedance tapers were fabricated in coplanar-waveguide structures, where the ground plane is not shown in the sketch. The SNSPI was biased at a constant current $I_B$ and read out by room-temperature amplifiers (amp.). When a photon was absorbed, two pulses propagated towards the nanowire's two ends with delays $\tau_1$ and $\tau_2$, from which the photon landing location and the photon arrival time were determined simultaneously. **b**, A scanning-electron-micrograph of the top nine rows (out of 15 rows in total) from a single 19.7-mm-long SNSPI that meandered over an active imaging area of 286 μm × 193 μm. The dimensions shown are $q = 13.0$ μm, $h = 9.7$ μm, and $p = 5.4$ μm. The scale bars in the inset figures are 2 μm (left) and 300 nm (right). **c**, A single-photon image of the pattern formed by light passing through a metal mesh, which was placed on top of the SNSPI with a gap of ~ 200 μm (see Methods). The circular periodic patterns reflect the opening holes of the mesh. The wavelength of the light was 780 nm. The image consists of data from 427,905 photon detection events. The color of the map shows the normalized photon counts at each location. **d**, An optical image of the metal mesh, whose opening size was 43 μm and whose wire diameter was 30 = μm. The scale bar is 50 μm.

To create a 2D image with horizontal and vertical addresses ($H,V$), the 1D nanowire is meandered to cover a 2D area. The spatial resolution in each direction could thus be optimized independently for particular applications (*e.g.*, near-field imaging at sub-wavelength resolution by reducing the spacing between adjacent sections of the nanowire). In our experiments, as shown in Fig. 1b, the nanowire was meandered in both the *H* direction and the *V* direction to have nearly equal spatial resolution in *H* and *V*. This double-meandered nanowire covered a rectangular area of 286 μm × 193 μm. To demonstrate the imaging process, we placed an object on top of this rectangular area and evenly illuminated photons through it to project a pattern on the nanowire (see Methods). Figure 1c shows an image of a metal mesh of a simple periodic structure whose opening holes correspond to the visible grid of circular patterns. Figure 2 shows the image construction formed by 989,897 photon detection events. A video (see supplementary online material) also demonstrates how the accumulation of detection events generated the image.

The spatial resolution of the imager was dominated by the electrical noise in the readout circuits and the speed of signal propagation in the transmission line. Consider a photon arriving at $x_p$ resulting in pulses observed at times $\tau_1$ and $\tau_2$. Electrical noise contributed to variation in the determination of $\tau_1$ and $\tau_2$, resulting in a variation of measured $x_p$ based on the function $x_p = ((\tau_2 - \tau_1)v + L)/2$. We qualified this uncertainty by defining a Gaussian point-spread function $b(x) = \exp(-x^2/2h^2)$ where $h = (\delta/\rho) \times v/2$ ($\delta$ is the amplitude of the electrical noise and $\rho$ is the slope of the pulses at the discrimination threshold level). The point-spread function can be used to estimate the effective resolution as limited by electrical noise. From the waveform of the output pulses, $(\delta/\rho)$ was determined to be 3.2 ps (see Methods). Given this constraint, a slow $v$ can help to reduce $h$ so that the original location of a photon-detection event can be determined with less error. For transmission lines, the velocity $v = 1/\sqrt{L_m C_m}$ (where $L_m$ and $C_m$ are the inductance and capacitance per unit length)[23]. In the SNSPI, as the thickness of superconducting nanowire is much less than the London penetration depth, the kinetic inductance of the nanowire is higher than the geometric inductance by about two orders of magnitude (see Methods for the calculation of kinetic inductance and geometric inductance)[21,26]. Additionally, we designed the gap between the ground and signal line to be 100 nm to increase $C_m$. As a result, the final $v$ = 5.6 μm/ps (~ 2% of the speed of light $c$). Substituting the measured value of $v$ into the point-spread function $b(x)$, we have $h = 8.9$ μm and the full-width-at-half-magnitude $f_w$ of $b(x)$ is 21 μm. The width of $b(x)$ determines the noise-limited spatial resolution, which will be compared with the measured width of the sharp peaks generated from dark counts in later discussion.

With the given point-spread function, the 2D spatial resolution can be calculated by taking into account the meander geometry, from which the 1D distance $x$ is mapped to the 2D location ($H,V$). In the geometry shown in Fig 1b, the vertical spatial resolution is the spacing between rows $q = 13.0$ μm and the horizontal spatial resolution is $f_w \times p/l_m = 5.6$ μm, where $l_m = 22.84$ μm is the effective length of one meander period in each row. With these spatial resolutions, we were able to image letters with a 12.6 μm stroke width and 12.6 μm spacing between strokes (see Fig. 2d). A smaller $h$ could be achieved by reducing the timing uncertainty $j_e$ or slowing down $v$ even further by using other superconducting materials with higher kinetic inductance (*e.g.*, tungsten silicide), substrates with higher dielectric constant (e.g., $LaAlO_3$), or other transmission line structures (e.g., microstrip lines).

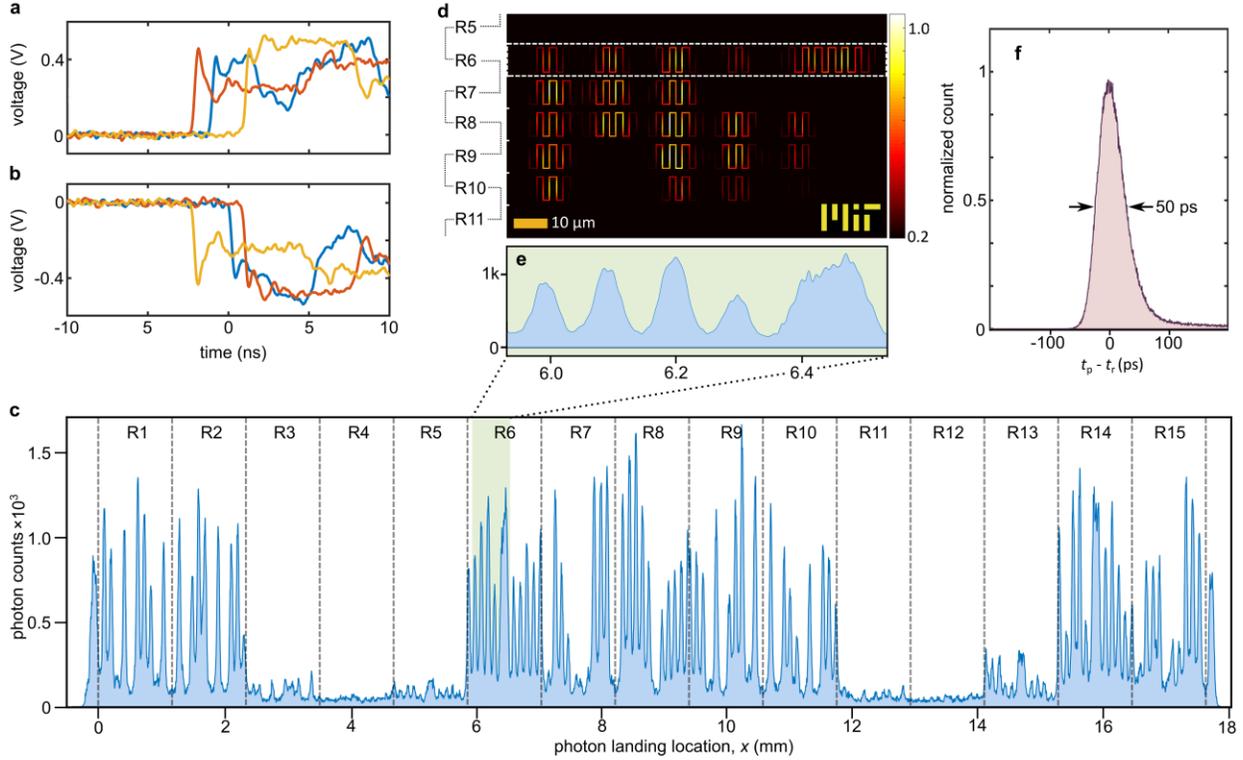

**Figure 2 | Spatial and temporal detections by the SNSPI. a**, **and b**, Oscilloscope waveforms of output pulses from the two ends of the nanowire from three photon detection events occurring at different locations. We illuminated the device with sub-ps optical pulses from a 1.5 μm mode-locked laser and determined $\tau_1$ and $\tau_2$ after removing the fixed delays outside of the nanowire from the measured arrival times of the pulses. **c**, Histogram of 989,897 photon detections displayed along the 1D distance coordinate $x$. We used a 405 nm continuous-wave light-emitting diode source to illuminate an array of rectangular shapes. We used the difference of pulse arrival times to determine photon position. The span of $x$ was divided into 15 sections corresponding to the 15 rows (R1~R15) of the double-meandered nanowire. **d**, Image of an institutional logo constructed from the photon-position data (only a portion of full image region is shown). **e**, Magnified section of histogram corresponding to the bright lines included within white box in row 6. **f**. Histogram of the difference between the photon arrival time $t_p$ measured by the SNSPI and a reference time $t_r$. We used a 1.5 μm mode-locked laser to illuminate the SNSPI and generate $t_r$ with ps resolution. The FWHM of the histogram profile was 50 ps, which was used for defining the timing jitter $j_d$.

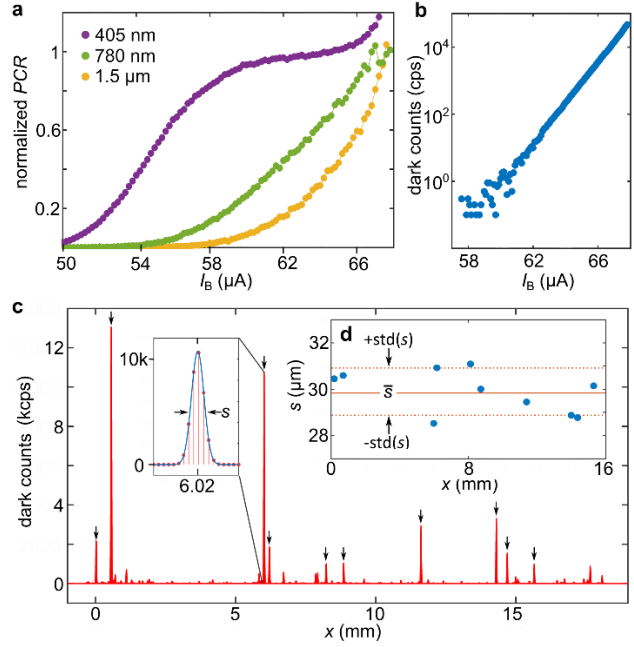

**Figure 3 | Detection performance of the SNSPI. a**, Normalized photon count rate (*PCR*) versus the bias current at wavelengths from visible to infrared. **b**, Overall dark counts versus bias current. **c**, Dark counts histogram over the 1D distance at a bias current of 65 μA. We selected ten peaks of highest amplitude (indicated by the arrows on top) and fit each of them with a Gaussian function, whose FWHM *s* is calculated. **d**, Distribution of *s* from the ten peaks. The average of *s* is $\bar{s}$ = 29.9 μm and the standard deviation is std(*s*) = 0.9 μm.

Photon arrival time $t_p$ was determined by using the equation $t_p = ((\tau_2 + \tau_1) - L/v)/2$, and was thus independent of the hot-spot location $x_p$. The temporal resolution was characterized by the timing jitter $j_d$ defined as the time variation of the measured photon arrival times, which included both the electronic jitter from noise and intrinsic jitter from the photon detection mechanism. To precisely measure $j_d$, we used a 1.5 μm mode-locked laser, whose output was split into two beams. One beam was sent to the SNSPI for measuring $t_p$ while the other beam was measured by a fast photodiode to determine a timing reference $t_r$ with ps resolution. Figure 2f shows the histogram of $t_p - t_r$. The FWHM of the histogram profile gave the timing jitter, which is $j_d$ = 50 ps. This value is consistent with reported timing jitters for SNSPDs and is significantly lower than the timing jitters of TESs and MKIDs. Aside from the high temporal resolution, the SNSPI also exhibited the wide optical bandwidth typical of conventional SNSPDs[17]. Figure 3a shows the photon counts versus bias current at wavelengths of 405 nm, 780 nm and 1.5 μm. At 405 nm wavelength, the internal quantum efficiency of the wire saturated, suggesting a near-unity internal quantum efficiency of the nanowire. At longer wavelengths, the quantum efficiency was reduced, as would be expected for wires that are 300 nm wide. Reducing the wire width, using microstrip transmission lines to remove the adjacent insensitive ground regions, and integrating an optical cavity with the device (which could use the microstrip ground plane as a reflecting mirror) is expected to increase the device efficiency in future application of this device.

A spatial map of dark counts of the SNSPI can be created by operating the device in a well-shielded environment with no illumination. As shown in Fig. 3b, the total number of dark count events summed over the length of the SNSPI exhibited an exponential dependence on bias current, similar to what is measured in SNSPDs. We took images with the SNSPI at a bias current of 60 μA, where the overall dark count rate was ~ 1 cps, ensuring a high signal-to-noise ratio of the image. To investigate the distribution of dark counts along the nanowire, we increased the bias current to 65 μA to have more dark counts and thus to reduce the overall acquisition time. Figure 3c shows the dark count histogram observed along the nanowire. We then calculated the FWHM $s$ for each of the ten peaks with the highest amplitudes. The average value of $s$ was $\bar{s} = 30$ μm and the variation (defined as the standard deviation) was $\mathrm{std}(s) = 0.9$ μm. Those ten peaks contributed to 74% of the total dark counts, while their combined length was only ~ 2% of the total length of the nanowire, indicating that the image quality was robust against dark counts. Dark counts could also be partially subtracted from the imager based on calibration from zero-light experiments.

The widths of the dark-count peaks were slightly larger than the spatial resolution calculated from the point-spread function, which suggests a possible intrinsic length of dark-count locations, or perhaps an underestimate of the system electrical noise. To estimate the number of resolvable locations in the SNSPI, we used the measured $\bar{s}$ as the minimum length that the SNSPI was able to distinguish. With this assumption, the maximum resolvable number of pixels $N_p$ in the SNSPI is $N_p = L_e/\bar{s} \cong 590$, where $L_e = 17.635$ mm is the effective length of a straight wire converted from the double-meandered geometry, taking into account the increase of signals velocity in corners. The estimated pixel density was thus ~ $10^6$ pixels per cm$^2$.

In summary, we have introduced a scalable architecture for single-photon imagers. By engineering the microwave behavior of a photon-sensitive nanowire into a plasmonic transmission line with a surprisingly slow velocity of 2% of the speed of light, we have demonstrated ~ 590 effective pixels, sub-20 μm spatial resolution and 50 ps temporal resolution in a 19 mm long superconducting nanowire. To implement SNSPIs into future applications, the microwave design of the nanowire can be combined with optical structures, such as an optical cavity, to match the device detection efficiency of state-of-the-art SNSPDs[2]. Future work may also exploit modeling and signal processing to extract photon number information from the device. Due to the reduced requirements for readout lines, by integrating many SNSPIs into a super-array, we envision a camera with millions of pixels, 10 GHz counting rate, and 100 cm$^2$ detection area. In addition, the image quality can be further enhanced through improved optics used to focus the object image and superior electronics with less noise and faster acquisition speed. With these improvements, we expect the SNSPI will be used as a powerful single-photon detection tool in various fields, such as quantum information science, communications, single-photon imaging and spectroscopy, and astronomical observation.

**Acknowledgements**

The authors thank Richard Hobbs, Chung-Soo Kim and Mark Mondol for their technical support in nanofabrication, and Philip Mauskopf, Joel K.W. Yang, Zheshen Zhang, and Emily Toomey for scientific discussion. This research was supported by the National Science Foundation (NSF) grants under contact No. ECCS-1509486 (MIT) and No. ECCS-1509253 (UNF), and the Air Force Office of Scientific Research (AFOSR) grant under contract No. FA9550-14-1-0052. Di Zhu is supported by National Science Scholarship from A*STAR, Singapore. Niccolò Calandri would like to thank his financial support from the Roberto Rocca project during his visit in MIT. Andrew Dane was supported by NASA Space Technology Research Fellowship (award number NNX14AL48H). Adam McCaughan was supported by a fellowship from the NSF iQuISE program (award number 0801525).

**Author Contributions**

Q.-Y. Z. and K. B. came with the initial idea. Q.-Y. Z. designed and fabricated the nanowire imager. Q.-Y. Z. and D. Z. took the optical measurements. Q.-Y. Z., N.C., F.B., and H.-Z. W. characterized initial devices. A. D. supported the superconducting films. Q.-Y. Z., A.M., and D. S. did the microwave simulations. Q.-Y. Z. analyzed the data and programmed the imaging script. K. B. supervised the project. Q.-Y. Z. and K. B. wrote the paper with input from all other authors.

# Methods

**Device fabrication and imaging pattern preparation**

*Device fabrication:* The SNSPI was fabricated from a ~7 nm thick niobium nitride (NbN) thin film. The film was deposited on a 4-inch silicon wafer with a 300-nm-thick surface layer of silicon dioxide. The NbN had a critical temperature of $T_C = 10K$, a sheet resistance of $R_s = 331\ \Omega$/square, and a residual resistance ratio of $RRR = 0.8$. The kinetic inductance was $L_K = h \cdot R_S/(2 \cdot \pi^2 \cdot \Delta \cdot \tanh\left(\frac{\Delta}{2 \cdot k_B \cdot T_c}\right)) = 49$ pH/square, where $\Delta = 1.76 \cdot k_B \cdot T_c \cdot \tanh\left(1.74 \cdot \sqrt{\frac{T_c}{T} - 1}\right)$ is the temperature-dependent superconducting energy gap, $h$ is the Planck constant, $k_B$ is the Boltzmann constant, and $T_c = 4.2$ K is the operating temperature[27]. The structure of the nanowire transmission line and the tapers were formed by using a 125 kV electron beam lithography tool and then transferred into the NbN layer in a $CF_4$ atmosphere using a reactive ion-etcher (more fabrication details are reported in previous publications of SNSPDs[28,29])

Two objects were imaged with the SNSPI. The object used to generate Fig. 1c was a stainless steel mesh (McMaster-Carr part number 34735K69) with an opening size of 43 μm and a wire diameter of 30 μm. The structure of the MIT logo image (Fig. 2d) was patterned by using photo-lithography from a lift-off process on an indium-tin-oxide (ITO) substrate. The letters were made from a bilayer of metals (50 nm thick chromium and 50 nm thick gold). For each measurement, we placed the object on top of the device, leaving a gap of about 200 μm between the mask and the device.

*Imaging Set-up*: Here, we give the details of the electrical readouts and the optical setup for operating the SNSPI for taking images, which is shown in supplementary Fig. 1. We mounted the SNSPI and the imaged pattern on a PCB board with two coaxial cables for connecting the device to room-temperature readout electronics. The PCB board was mounted on a 2-inch lens tube which has a lens for collimating the laser beam. The SNSPI, imaged pattern, PCB and the tube were immersed in liquid helium at 4.2 K.

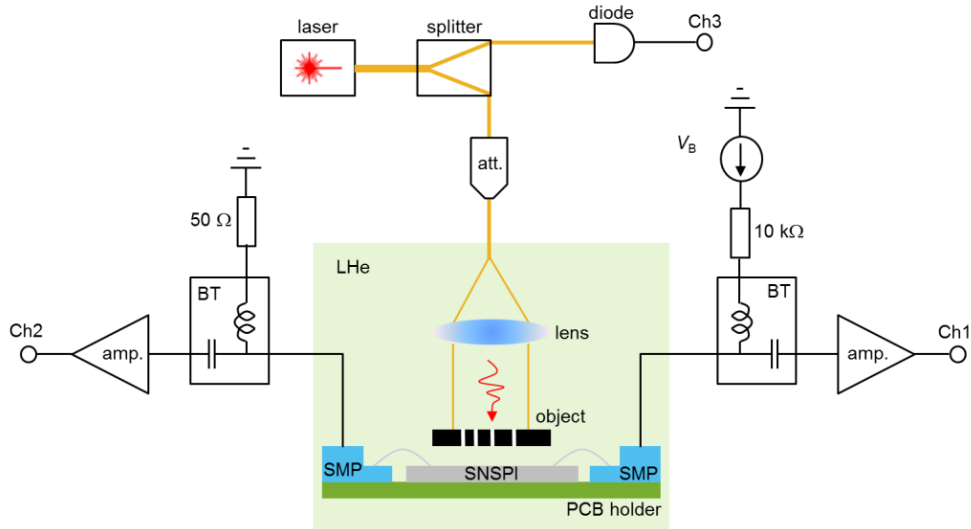

**Supplementary Figure 1 | Detailed experimental set-up** A detailed description is in the Method section 'setup'. The abbreviations are: att. (optical attenuation), $V_B$ (bias voltage source), BT (bias tee), amp (RF amplifier), SMP (SMP connector), Ch1~Ch3 (the input channels of an oscilloscope), LHe (liquid helium).

The room-temperature readout circuit was used for supplying the SNSPI with a DC bias current, amplifying the output pulses, and determining the pulse arrival times. We used two bias tees (Mini Circuits ZFBT-6GW+) to separate the DC path and the RF path. As the SNSPI is a two-terminal device, we terminated one bias tee's DC port with a 50 Ω load and connected a voltage source (SRS 928) in series with a 10 kΩ resistor to the other bias tee's DC port. With this configuration, a DC current could go through the nanowire. Each bias tee's RF ports were connected to two RF amplifiers in series (RFbay LNA-2500 in serial with Mini Circuit ZX60-3018G-S), offering a total gain of 47 dB and a bandwidth of 20 MHz~2.5 GHz. The two outputs were acquired by a 6 GHz real-time oscilloscope (Lecroy Wavepro 760Zi). The oscilloscope was triggered by one of the output channels. The arrival times were recorded in the oscilloscope and then exported to computers for image post processing. To detect the arrival times of the electrical pulse, we set a constant discrimination threshold level on the pulse edges. Due to the existence of voltage noise that changed the amplitude of the pulse, the measured arrival times contained another timing variation, which was defined as the electrical jitter $j_e$. We calculated $j_e = \delta_n/\rho$ from the measured waveforms of the pulses, where $\delta_n$ was the root-mean-square noise in a 20 ns timing duration and $\rho$ was the slew rate at the discrimination threshold level. In our setup, for each channel, $j_e$ was $3.2 \pm 0.3$ ps.

The light was prepared at room temperature and then illuminated on the device through a single-mode fiber (SMF-28) connecting to the lens tube. We used three light sources at different wavelengths, 405 nm, 780nm and 1.5 μm. The light was attenuated to ensure the device was triggered by a single-photon. The 1.5 μm source was a mode-locked laser (Calmar Laser FPL-02CFF), offering sub-ps wide optical pulses at a repetition rate of 20 MHz. We used this source for characterizing the temporal resolution of the SNSPI. In this measurement, the output of the laser was split into two: one illuminated the imager and the other one was detected by a fast photodiode (Thorlabs DET08CFC) as a timing reference. The 780 nm light was a continuous-wave laser source (Thorlabs S1FC780), and the 405 nm light was a fiber-coupled continuous-wave light-emitting diode (Thorlabs M405F1). When the continuous-wave sources were used, the splitter and photodiode were removed.

**Microwave design of the SNSPI**

*Superconducting kinetic inductance plasmonic transmission line*: We calculated the impedance and velocity of a superconducting transmission line both numerically and analytically. In the analytical calculation, we first calculate the capacitance and inductance of a convention coplanar waveguide (CPW) made from lossless metal and then replaced the inductance with the total of the kinetic inductance and the geometrical inductance. As shown in supplementary Fig. 2a, both methods showed similar values.

*Wideband impedance transformation by using taper structures*: We used the Klopfenstein taper for transforming the nanowire impedance to 50 Ω to preserve the fast rising edge of a photon detection pulse. In order to verify the taper's performance, we fabricated a 17 mm long NbN taper without a photon-sensitive nanowire at the middle. The taper was designed into a CPW structure with a fixed 3 μm gap to the ground plane and a signal line whose width smoothly changed from 88 μm at the two ends to 10 μm in the center. In order to characterize the superconducting taper without switching it to the normal by the input signals, the narrowest width of the nanowire in the center was designed to 10 μm to have a switching current of 0.4 mA.

The bandwidth of the taper was measured by a Network analyzer (see supplementary Fig. 2b). The pass band of the taper started at 0.7 GHz, which was able to cover the spectrum of the fast edge triggered by photon absorption. The performance of the taper also validated the calculation of the superconducting transmission line. Although the pass band stopped at 2.4 GHz due to the loss of the PCB board and the bonding wires, the bandwidth was sufficient to support the read of the fast output pulses.

We demonstrated the pulse propagation by sending a pulse into the taper and measuring the output pulse from the other end. A 200-ps-wide electrical pulse (Avtech, AVMP-2-C-P-EPIA) was split into two. One pulse was acquired by a 6 GHz oscilloscope as a timing reference, while the other pulse was fed into the taper and its transmitted signal was acquired by another channel of the oscilloscope. We also compared the transmitted signal from the taper to the transmitted signal from a 50 Ω transmission line with the same length (corresponding to a delay of 94 ps) but made from a PCB. The delay difference between a PCB transmission line and the taper was 760 ps, indicating the superconducting taper slowed down the average velocity to 11% of a PCB transmission line. The amplitude of the transmitted signal from a taper reduced to 60% of the transmitted signal from the PCB transmission line; however, the rising edge of the pulse was well preserved, verifying that the taper helped the propagation of the fast pulse through a wire of mismatched impedance.

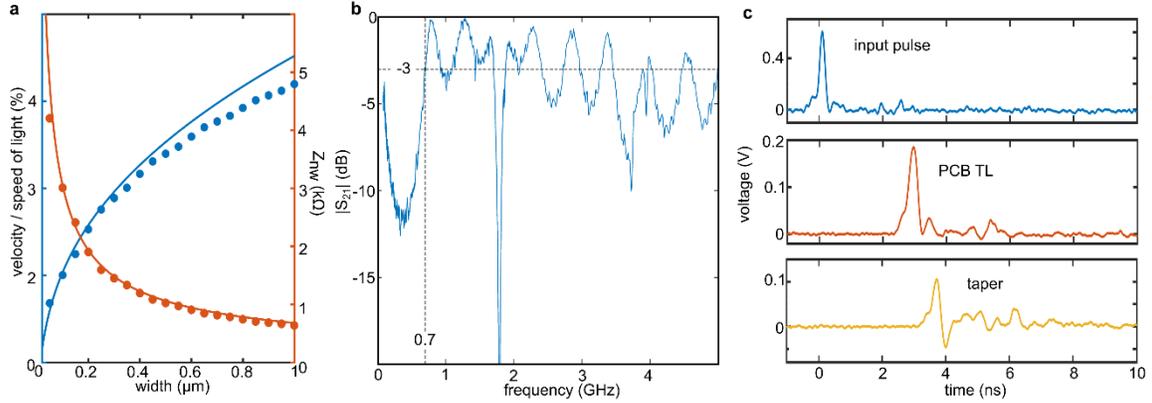

**Supplementary Figure 2 | Microwave design of the SNSPI a**, Calculated velocity and impedance of a superconducting coplanar waveguide at different widths. The dots are from numerical simulation while the lines are from analytical calculation. See Methods for details about the geometry of the wire. **b**, Magnitude of the transmission coefficient $S_{21} = 20 \times \log(V_{out}/V_{in})$ of a 17.3-mm-long straight taper without connecting to a photon-sensitive nanowire, where $V_{in}$ is the input signal to the taper from a network analyzer and $V_{out}$ is the output signal through the taper. The high-pass bandwidth starts at 0.7 GHz. The sharp dips close to 2 GHz and 4 GHz are probably from the wire bonding connections between the chip and the print circuit board (PCB). **c**, Time domain measurement of the taper. We measured the transmitted pulse propagating through the taper and compared it to the transmitted pulse propagating through a 50 Ω PCB transmission line (TL) for evaluating the amplitude loss and the pulse shape distortion of the transmitted pulse propagating through the superconducting taper.

In the SNSPI used for taking images, the taper on each end was designed to have a bandwidth, whose band pass started at 0.8 GHz. Each taper has an overall length of 27 mm, with its width smoothly changing from 105 μm to 300 nm. As shown in Fig. 2a, the rising edge of the photon detection pulses was about 240 ps without any reflection, ensuring the precise measurements of delay times $\tau_1$ and $\tau_2$.

**Imaging processing algorithm**

The raw imaging data were derived from the histogram of differential time $\Delta t = \tau_2 - \tau_1$ of photon counts acquired by the oscilloscope. The purpose of the image processing was to map the photon count $C_n$ at each time bin $\Delta t_n$ to the intensity $I_n$ at the corresponding 2D locations $(H_n, V_n)$. First, we calculated the effective 1D distance $x$ from the layout of the double-meandered nanowire, taking into account the effective length of the corner (the corner's effective length was chosen to be 0.68 of its physical length based on numerical simulation of the propagation time through a corner) so that we determined a look-up table mapping $x_n$ to $(H_n, V_n)$. Secondly, we interpolated the histogram data with a finer time step of 0.045 ps and then converted $\Delta t_n$ to $x_n$ by using the formula $x_n = \frac{\Delta t_n + L}{2}$, where $v = 5.56$ μm/ps was the velocity and $L = 17.635$ mm was the effective length. Thus, we were able to map the time bin $t_n$ for each photon detection number $C_n$ to the 1D distance $x_n$. Finally, we set a 2D image frame with grid sizes of 0.5 μm in both directions, where the 2D locations $(H_n, V_n)$ were sorted. For each $(H_n, V_n)$, the intensity was set to the photon count $C_n$ whose corresponding time bin was $t_n$ and the 1D distance was $x_n$. To reduce the mapping time, we spread the photon count $C_n$ along the meandered nanowire at $(H_n, V_n)$ in the 2D image, where the photon counts were distributed according to a Gaussian weight function with a standard deviation of 5 μm. During the mapping process, we shifted $t_n$ with a constant time to correct for the difference of the delays from electrical connections to the two ends of the wire, and adjusted the velocity to result in a better image quality. The success of the tuning was evaluated by checking the alignments of neighboring rows.

Instead of using the histogram data, the image can be also constructed from compilation of single-photon detection events to offer a real-time single-photon video. To use the same image algorithm discussed previously, raw data of $(\Delta t, 1)$ was used, where each time difference was assigned to one photon count. The Supplementary Video shows a demonstration of such video by taking the original differential arrival times from the oscilloscope.

**Maximum counting rate of the SNSPI**

The imaging time was limited by the counting rate of the SNSPI and the acquisition speed of the readout. In our present setup, we used a 6 GHz real-time oscilloscope because of its sub-ps intrinsic timing jitter. However, the oscilloscope had a refresh rate of ~100 counts per second, which was the bottleneck that limited the overall imaging time. To investigate the ultimate speed of the SNSPI, we measured the maximum counting rate ($CR_{max}$) of the SNSPI, which was defined as the count rate when we found that the average detection efficiency dropped by a half. As shown in supplementary Fig. 3, $CR_{max}$, was 2 cps measured for 1.5 µm photons. The measured $CR_{max}$ indicated the imaging time for an image such as the one shown in Fig.1c or Fig. 2d would require a few seconds if the oscilloscope bottleneck were removed. The measured $CR_{max}$ was lower than typical SNSPDs because it had a higher total kinetic inductance (estimated to be ~ 3 µH due to the long wire and the tapers), increasing the time for the bias current in the nanowire to recover[15].

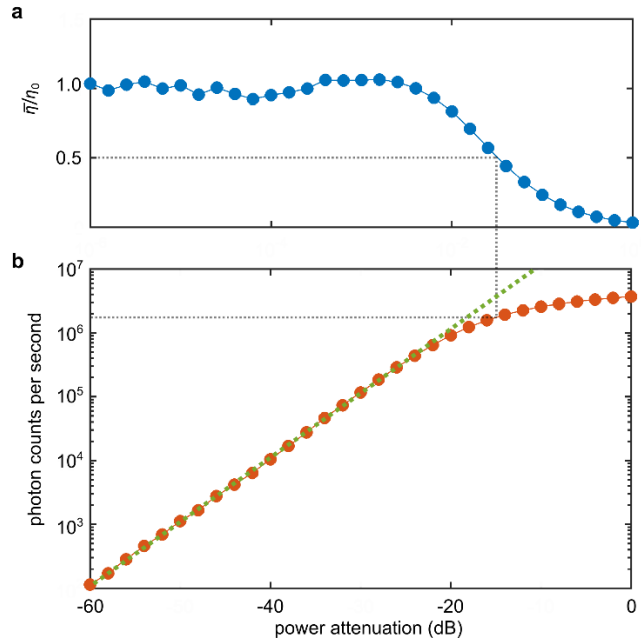

**Supplementary Figure 3 | Maximum counting rate of the SNSPI a**, The average detection efficiency ($\bar{\eta}$) per second normalized to the detection efficiency for detecting a single photon ($\eta_0$) at different attenuations of the incident light. We used a 1.5 μm continuous-wave laser and the readout circuit reported in ref. [30] to avoid the capacitive charging from RF amplifiers. **b**, The photon counts per second versus the attenuation of the incident light. The linear part, shown as the green line, gives $\eta_0$, where the inter arrival time of individual photons is longer than the current recovery time.

**References in Methods section**